\documentstyle[a4,12pt,epsf,amsfonts,amssymb]{article}  

\DeclareFontFamily{U}{rsfs}{\skewchar\font"7F} 
\DeclareFontShape{U}{rsfs}{m}{n}{<-6> rsfs5 <6-8> rsfs7 <8-> rsfs10}{} 
\DeclareMathAlphabet{\mathscr}{U}{rsfs}{m}{n}  

\def\ben{\begin{equation}} 
\def\een{\end{equation}}
\def\bena{\begin{eqnarray}} 
\def\eena{\end{eqnarray}}
\def\f(#1/#2){\frac{#1}{#2}} 
\def\Frac(#1/#2){\left(\frac{#1}{#2}\right)}

\begin{document} 

\title{ The History and Present Status of Quantum Field Theory in Curved
Spacetime}  

\author{Robert M. Wald \\ \\  
 {\it Enrico Fermi Institute and Department of Physics}\\  
 {\it The University of Chicago, Chicago, IL 60637, USA} 
        } %%% 

\maketitle

\begin{abstract}

Quantum field theory in curved spacetime is a theory wherein matter is
treated fully in accord with the principles of quantum field theory,
but gravity is treated classically in accord with general
relativity. It is not expected to be an exact theory of nature, but it
should provide a good approximate description in circumstances where
the quantum effects of gravity itself do not play a dominant role.
Some of the earliest applications of the theory were to
study particle creation effects in an expanding universe. A major
impetus to the theory was provided by Hawking's calculation of
particle creation by black holes, showing that black holes radiate as
perfect black bodies. During the past 30 years, considerable progress
has been made in giving a mathematically rigorous formulation of
quantum field theory in curved spacetime. Major issues of principle
with regard to the formulation of the theory arise from the lack of
Poincare symmetry, the absence of a preferred vacuum state, and, in
general, the absence of asymptotic regions in which particle states
can be defined. By the mid-1980's, it was understood how all of these
difficulties could be overcome for free (i.e., non-self-interacting)
quantum fields by formulating the theory via the algebraic approach
and focusing attention on the local field observables rather than a notion
of ``particles''. However, these
ideas, by themselves, were not adequate for the formulation of
interacting quantum field theory, even at a perturbative level, since
standard renormalization prescriptions in Minkowski spacetime rely
heavily on Poincare invariance and the existence of a Poincare
invariant vacuum state. However, during the past decade, great
progress has been made, mainly due to the importation into the theory
of the methods of ``microlocal analysis''. This article will describe
the historical development of the subject and describe some of the
recent progress.

\end{abstract} 

\section{Introduction} 

Quantum field theory in curved spacetime is the theory of quantum
fields propagating in a classical curved spacetime. 
Here the spacetime
is described, in accord with general relativity, by a manifold, $M$,
on which is defined a Lorentz metric, $g_{ab}$. In order to assure
that classical dynamics is well defined on $(M, g_{ab})$, we restrict
attention to the case where $(M, g_{ab})$ is globally hyperbolic (see,
e.g., \cite{w1}). In the framework of quantum field theory in curved
spacetime, back-reaction of the quantum fields on the spacetime
geometry can be taken into account by imposing the semi-classical
Einstein equation $G_{ab} = 8 \pi \langle T_{ab} \rangle$. However, I will not
consider issues associated with back-reaction here, so in the
following, $(M, g_{ab})$ may be taken to be an arbitrary, fixed
globally hyperbolic spacetime.

This article will focus primarily on issues concerning the formulation
of quantum field theory in curved spacetime. I will describe some
aspects of the historical evolution of the subject through the
mid-1970's, at which point it had become clear that a proper
formulation of the theory could not be based upon a notion of
``particles''. I will then describe how the major conceptual obstacles
to formulating the theory in the case of free quantum fields were
overcome by adopting the algebraic approach. Finally, I will describe some
of the progress that has been made during the past decade toward the
formulation of interacting quantum field theory in curved spacetime.

Much of the quantum theory of a free field follows directly from the
analysis of an ordinary quantum mechanical harmonic oscillator,
described by the Hamiltonian
\begin{equation}
H = \frac{1}{2} \, p^2 + \frac{1}{2} \omega^2 \, q^2 \, .
\end{equation}
By introducing the ``lowering'' (or ``annihilation'') operator
\begin{equation}
a \equiv \sqrt{\frac{\omega}{2}} \,\, q + i \sqrt{\frac{1}{2\omega}} \,\, p
\end{equation}
we can rewrite $H$ as
\begin{equation}
H = \omega (a^\dagger a + \frac{1}{2} I)
\end{equation}
where $a^\dagger$ is referred to as the ``raising'' (or ``creation'')
operator, and we have the commutation relations 
\begin{equation}
[a, a^\dagger] = I \, , \,\,\, [H,a] = - \omega a \, .
\end{equation}
It then follows that in the Heisenberg representation, the position
operator, $q_H$, is given by
\begin{equation}
q_H = \sqrt{\frac{1}{2\omega}} \left(e^{-i\omega t} a 
+ e^{i\omega t} a^\dagger \right)
\end{equation}
so that $a$ is seen to be the positive frequency part of the
Heisenberg position operator. The ground state, $|0 \rangle$, of the harmonic
oscillator is determined by
\begin{equation}
a |0 \rangle = 0 \, .
\end{equation}
All other states of the harmonic oscillator are obtained by successive
applications of $a^\dagger$ to $|0 \rangle$.

Consider, now, a free Klein-Gordon scalar field, $\phi$, in Minkowski
spacetime. Classically, $\phi$ satisfies the wave equation
\begin{equation}
\partial^a \partial_a \phi - m^2 \phi = 0 \, .
\end{equation}
To avoid technical awkwardness, it is convenient to imagine that the
scalar field resides in a cubic box of side $L$ with periodic boundary
conditions. In that case, $\phi(t, \vec{x})$ can be decomposed in
terms of a Fourier series in $\vec{x}$. In terms of the Fourier coefficients
\begin{equation}
\phi_{\vec{k}} \equiv L^{-3/2} \int 
e^{-i\vec{k} \cdot \vec{x}} \phi(t, \vec{x}) \,\, d^3 x
\end{equation}
where 
\begin{equation}
\vec{k} = \frac{2 \pi}{L} (n_1, n_2, n_3)
\end{equation}
we have
\begin{equation}
H = \sum_{\vec{k}} \frac{1}{2} \left(|\dot{\phi}_{\vec{k}}|^2 
+ \omega^2_{\vec{k}} |\phi_{\vec{k}}|^2 \right)
\end{equation}
where
\begin{equation}
\omega^2_{\vec{k}} = |\vec{k}|^2 + m^2 \, .
\end{equation}

Thus, a free Klein-Gordon field, $\phi$, is seen to be nothing more
than an infinite collection of decoupled harmonic oscillators. The
quantum field theory associated to $\phi$ can therefore be obtained by
quantizing each of these oscillators. It follows immediately that the
Heisenberg field operator $\phi (t, \vec{x})$ should be given by the formula
\begin{equation}
\phi (t, \vec{x}) = L^{-3/2} \sum_{\vec{k}} \frac{1}{2 \omega_{\vec{k}}}
\left(e^{i \vec{k} \cdot \vec{x} -i \omega_{\vec{k}} t} a_{\vec{k}} 
+ e^{-i \vec{k} \cdot \vec{x} + i \omega_{\vec{k}} t} a^\dagger_{\vec{k}}  
\right) \, .
\label{phi}
\end{equation}
However, the sum on the right side of this equation does not converge
in any sense that would allow one to define the operator $\phi$ at the
point $(t, \vec{x})$. Roughly speaking, the infinite number of arbitrarily
high frequency oscillators fluctuate too much to allow $\phi (t,
\vec{x})$ to be defined. However, this difficulty can be overcome by
``smearing'' $\phi$ with an arbitrary ``test function'', $f$ (i.e., 
$f$ is a smooth
function of compact support), so as to define 
\begin{equation}
\phi(f) = \int f(t, \vec{x}) \phi (t, \vec{x}) d^4x 
\end{equation}
rather than $\phi (t, \vec{x})$. The resulting formula 
for $\phi(f)$ can be shown to make
rigorous mathematical sense, thus defining $\phi$ as an
``operator-valued distribution''.

The ground state, $|0 \rangle$, of $\phi$ is simply simultaneous ground state of all
of the harmonic oscillators that comprise $\phi$, i.e., it is the
state satisfying $a_{\vec{k}} |0 \rangle= 0$ for all $\vec{k}$. In quantum
field theory, this state is interpreted as representing the
``vacuum''. A state of the form $(a^\dagger)^n |0 \rangle$ is interpreted as
a state where a total of $n$ particles are present. In an interacting
theory, the state of the field may be such that the field behaves like
a free field at early and late times. In that case, we would have a
particle interpretation of the states of the field at early and late
times. The relationship between the early and late time particle
descriptions of a state---given by the S-matrix---contains a great
deal of the dynamical information about the interacting theory, and,
indeed, contains all of the information relevant to laboratory
scattering experiments.

The particle interpretation/description of quantum field theory in
flat spacetime has been remarkably successful---to the extent that one
might easily get the impression from the way the theory is normally
described that, at a fundamental level, quantum field theory is really
a theory of particles. However, the definition of particles relies on
the decomposition of $\phi$ into annihilation and creation operators
in eq.(\ref{phi}). This decomposition, in turn, relies heavily on the
time translation symmetry of Minkowski spacetime, since the
``annihilation part'' of $\phi$ is its positive frequency part with
respect to time translations. In a curved spacetime that does not
possess a time translation symmetry, it is far from obvious how a
notion of ``particles'' should be defined.

\section{The Development of Quantum Field Theory in Curved Spacetime 
from the mid-1960's Through the mid-1970's}

Beginning in the mid-1960's, Parker investigated effects of
particle creation in an expanding universe \cite{p1,p2}. Consider a
spatially flat Friedmann-Lemaitre-Robertson-Walker spacetime, with metric
\begin{equation}
ds^2 = - dt^2 + a^2(t) [dx^2 + dy^2 + dz^2] \, .
\end{equation}
Consider, first the (highly artificial) case where $a(t)$ is constant
for $t<t_0$ and is again constant for $t>t_1$, but goes through a
time-dependent phase at intermediate times, $t_0 \leq t \leq t_1$. In
the ``in'' region $t<t_0$, spacetime is locally indistinguishable from
a corresponding portion of Minkowski spacetime, so a given state of a
free quantum field will have a particle interpretation in that region,
i.e., it can be characterized by its ``particle content''. Similarly,
in the ``out'' region, $t>t_1$, the same state will also have a
particle interpretation. However, on account of the time dependence of
the metric in the intermediate region, a classical solution of the
Klein-Gordon equation that corresponds to a purely positive frequency
solution in the ``in'' region will not correspond to a purely positive
frequency solution in the ``out region''. This means that the ``in''
and ``out'' annihilation and creation operators of the quantum field
(corresponding to the decomposition of eq.(\ref{phi}) for the ``in''
and ``out'' regions) will be different. This, in turn, implies that
the particle content in the ``in'' and ``out'' regions will be
different. In other words, the expansion of the universe will result
in spontaneous particle creation.  Quite generally, the relationship
between the ``in'' and ``out'' annihilation and creation operators is
given by a {\it Bogoliubov transformation}, whose coefficients are
determined by the classical scattering. It is not difficult to derive
a general formula for the resulting S-matrix in terms of these
coefficients and, in particular, an expression for the spontaneous
particle creation from the vacuum (see, e.g., \cite{w2} for a general
derivation).

Of course, we do not believe that the universe began or will end in a
phase where $a(t)$ is constant. How do we analyze and describe
particle creation in a more realistic context? If the universe is
expanding sufficiently slowly, an approximate notion of an ``adiabatic
vacuum state'' can be defined \cite{p2}, and a notion of ``particles''
relative to this adiabatic vacuum can be introduced. Such a notion of
particles is completely adequate in the present universe to describe
quantum field phenomena on scales small compared with the Hubble
radius, i.e., only when we consider modes of the field whose
oscillation period is comparable to or larger than the Hubble time
(i.e., $10^{10}$ years in the present universe) does the notion of
``particles'' become genuinely ambiguous.  However, at or very near
the ``big bang'' singularity, the notion of ``particles'' is highly
ambiguous.

The next major steps in the development of quantum field theory in
curved spacetime came from the application of the theory to black
holes.  By definition, a {\it black hole} in an asymptotically flat
spacetime is a region of spacetime from which nothing can escape to
infinity. A black holes is believed to be the endpoint of the complete
gravitational collapse of a body.  The ``time reverse'' of a black
hole---i.e., a region of spacetime that is impossible to enter if one
starts from infinity---is called a {\it white hole}. It is believed
that white holes cannot occur in nature. (The asymmetry between the
expected occurrence of black holes and the expected non-occurrance of white
holes in nature is undoubtedly closely related to the second law of
thermodynamics.)  However, black holes are expected to ``settle down''
to a stationary final state, and if one extends the idealized
stationary final state metric of a black hole backward in time
preserving the stationary symmetry, one obtains a spacetime containing a
white hole region. Thus, although white holes are not expected to
occur in nature, they do occur in the mathematically idealized
solutions used to describe the final stationary states of black holes.

Since, by definition, nothing can escape from a black hole, it would
seem that black holes would be one of the least promising places to
seek any observable effects of particle creation. However, the study
of effects of particle creation by black holes arose quite naturally
for reasons that I shall now explain. Outside of a rotating black
hole is a region, called the {\it ergosphere}, where the Killing field
that describes time translations at infinity becomes spacelike. This
means that an observer in the ergosphere cannot ``stand still''
relative to a stationary observer at infinity. In fact, an observer in
the ergosphere must rotate relative to infinity in the same direction
as the rotation of the black hole; this is an extreme example of the
``dragging of inertial frames'' effect in general relativity. The
prime importance of the ergosphere is that, since the time translation
Killing field is spacelike there, it is possible to have classical
particles whose total energy (including rest mass energy) relative to
infinity is negative. Consequently, as Penrose realized in 1969, one
can extract energy from a black hole by sending a body into the
ergosphere and having it break up into two fragments, one of which has
negative total energy. The negative energy fragment then falls into
the black hole (thereby reducing its mass), but it can be arranged
that the positive energy fragment emerges to infinity, carrying
greater total energy than the original body.

Not long after Penrose's discovery, Misner (unpublished) and
Zel'dovich and Starobinski \cite{z,s} realized that there is a wave
analog of the Penrose energy extraction process. Instead of sending in
a classical particle and having it break up into two fragments, one
can simply have a classical wave impinge upon a rotating black
hole. Part of this wave will be absorbed by the black hole and part
will return to infinity. However, if the frequency and angular
dependence of the wave are chosen to lie in the appropriate range,
then the part of the wave that is absorbed by the black hole will
carry negative energy relative to infinity. The portion of the wave
which returns to infinity will thereby have greater energy and
amplitude than the initial wave. This phenomenon is known as {\it
superradiant scattering}.

Thus, when a wave of superradiant frequency and angular dependence
impinges upon a rotating black hole, the black hole amplifies the wave
just like a laser. Superradiant scattering thus appears to be a direct
analog of stimulated emission. However, in quantum theory, it is well
known that in circumstances where stimulated emission occurs,
spontaneous emission will also occur. This suggested that for a
rotating black hole, ``spontaneous emission''---i.e., spontaneous
particle creation from the vacuum---should occur. This was noted by
Starobinski \cite{s} and confirmed by Unruh \cite{u1}.

The fact that spontaneous particle creation occurs near rotating black
holes did not cause much surprise or excitement. The effect is
negligibly small for macroscopic black holes such as those that would
be produced by the collapse of rotating stars, so unless tiny black
holes were produced in the early universe, the effect is not of
astrophysical importance. While it is an interesting phenomenon as a
matter of principle, it was not surprising or unexpected in view of
the ability to extract energy from a rotating black hole by classical
processes. However, it led directly to a further development that
caused a genuine revolution.

The calculation of particle creation by a rotating black hole was done
in the idealized spacetime representing the stationary final state of
the black hole. As explained above, this spacetime also contains a white
hole. Consequently, in the particle
creation calculation one has to impose initial conditions on the white
hole horizon that express the condition that no particles are emerging
from the white hole. In the calculation of Unruh \cite{u1}, a seemingly natural
choice of ``in'' vacuum state on the white hole horizon was made. But
it was not obvious that this choice was physically correct.

In 1974, Hawking \cite{h1} realized that this difficulty could be
overcome by considering the more physically relevant case of a
spacetime describing gravitational collapse to a black hole rather
than the idealized spacetime describing a stationary black hole (and
white hole). When he carried through the calculation, he found that
the results were significantly altered from the results obtained
for the idealized stationary black hole using the seemingly natural
choice of vacuum state on the white hole horizon. Remarkably, Hawking found
that even for a non-rotating black hole, particle creation occurs at
late times and produces a steady, non-zero flux of particles to
infinity. Even more remarkably, he found that, for a non-rotating
black hole, the spectrum of particles emitted to infinity 
at late times is precisely
thermal in character, at a temperature $T = \kappa/2 \pi$, where
$\kappa$ denotes the surface gravity of the black hole.

The implications of Hawking's result were enormous. It established
that black holes are perfect black bodies in the thermodynamic sense
at a non-zero temperature. This tied in beautifully with the
mathematical analogy that had previously been discovered between certain
laws of black hole physics and the ordinary laws of thermodynamics,
giving clear evidence that the similarity of these laws is much more
than a mere mathematical analogy. The identification of these
laws led to the identification
of $A/4$ as representing the physical entropy of a black hole, where
$A$ denotes the area of the event horizon. These and other
ramifications of Hawking's results have provided us with some of the
deepest insights we presently have regarding the nature of quantum gravity.

However, the Hawking calculation also had other major ramifications
for the development of quantum field theory in curved spacetime, and
these are the ones that I wish to emphasize here.  Although Hawking's results
were too beautiful to be disbelieved, there was a very disturbing
feature of the calculation: Using a seemingly natural notion of
``particles'' near the event horizon of the black hole, there appeared to
be a divergent density of ultra-high-frequency
particles present there. What do these ``particles'' mean? Does
their presence destroy the black hole?

To gain insight into this issue, Unruh \cite{u2} proceeded by taking a
purely operational viewpoint regarding the notion of ``particles'': A
``particle'' is a state of the field that makes a particle detector
register. Unruh then showed that in Minkowski spacetime, when a
quantum field is in it's ordinary vacuum state, a particle detector
carried by an accelerating observer will get excited. Indeed, he
showed that a uniformly accelerating observer ``sees'' an exactly
thermal spectrum of particles, at a temperature $T = a/2 \pi$, where
$a$ denotes the acceleration of the observer. This result provided an
explanation of the meaning of the divergent density of
ultra-high-frequency particles present near the event horizon of a
black hole. Such particles would be ``seen'' by a stationary observer
just outside the black hole. Indeed, such an observer would have to
undergo an enormous acceleration in order to remain stationary, and
what this observer sees corresponds exactly to the Unruh effect in
Minkowski spacetime. However, an observer who freely falls into the
black hole would not ``see'' these particles, just as an inertial
observer in Minkowski spacetime does not see the particles associated
with the accelerating observer. Furthermore, there are no significant
stress-energy effects associated with the quantum field near the
horizon of the black hole, so the presence of these ``particles'' as
seen by the stationary observer outside the black hole does not have a
significant back-reaction effect and, in particular, does not destroy
the black hole.

The clear lesson from Unruh's work is that one cannot view the notion
of ``particles'' as fundamental in quantum field theory. As its name
suggests, quantum field theory is truly the quantum theory of {\it
fields}, not particles. If one views the local fields as the
fundamental objects in the theory, the Unruh effect is seen to be a
simple consequence of how these fields interact with other quantum
mechanical systems (i.e., ``particle detectors''). If one attempts to view
``particles'' as the fundamental entities in the theory, the Unruh
effect becomes incomprehensible.

Furthermore, with the exception of stationary spacetimes (and certain
other spacetimes with very special properties), there is no preferred
notion of a ``vacuum state'' in quantum field theory in curved
spacetime and, correspondingly, there is no preferred notion of
``particles''. The difficulty is not that there is no notion of a
vacuum state but rather that there are many, and, in a general
spacetime, none can be uniquely singled out as having distinguished
properties. Thus, for this reason alone, it clearly would be
preferable to have a formulation of quantum field theory in curved
spacetime that does not require one to specify a vacuum state or a
notion of ``particles'' at the outset.

The usual way of constructing the theory of a free quantum field would
be to choose a vacuum state and then take the Hilbert space of states
to be the Fock space based upon this choice of vacuum state. The field
operator can then be defined (as an operator-valued-distribution) by
the analog of eq.(\ref{phi}). If different choices of vacuum state
corresponded to a mere relabeling of the states in terms of their
particle content, then it would make sense to construct the theory in
this manner, despite the lack of a preferred vacuum state. However, in
general, it turns out that different choices of vacuum state will give
rise to unitarily inequivalent theories, so the choice does
matter. Since there does not appear to be a preferred construction,
how does one formulate quantum field theory in a general curved
spacetime?

By the mid-1980's, it was well understood---via the efforts of
Ashtekar \cite{a}, Sewell \cite{se}, Kay \cite{k}, and others---that
the theory of a free quantum field in curved spacetime could be
formulated in an entirely satisfactory manner via the algebraic
approach. I shall now describe this formulation.

\section{The Algebraic Formulation of Free Quantum Field Theory 
in Curved Spacetime}

In the algebraic formulation of quantum field theory in an arbitrary
globally hyperbolic, curved spacetime, $(M, g_{ab})$, one begins by
specifying an algebra of field observables. For a free Klein-Gordon
field, a suitable algebra can be defined as follows. Start with the
free *-algebra, ${\mathcal A}_0$, generated by a unit element $I$ and
expressions of the form ``$\phi(f)$'', 
where $f$ is a test function on $M$. In other words,
${\mathcal A}_0$ consists of all formal finite linear combinations of
finite products of $\phi$'s and $\phi^*$'s, e.g., an example of an
element of ${\mathcal A}_0$ is $c_1 \phi(f_1) \phi(f_2) + c_2 \phi^*
(f_3) \phi(f_4) \phi^*(f_5)$. Now impose the following relations on
${\mathcal A}_0$: (i) linearity of $\phi(f)$ in $f$; (ii) reality of
$\phi$: $\phi^*(f) = \phi(\bar{f})$ where $\bar{f}$ denotes the
complex conjugate of $f$; (iii) the Klein-Gordon equation:
$\phi([\nabla^a \nabla_a - m^2] f) = 0$; (iv) the canonical
commutation relations:
\begin{equation}
[\phi(f), \phi(g)] = - i \Delta(f,g) I \, ,
\end{equation}
where $\Delta$ denotes the advanced minus retarded Green's
function. The desired *-algebra, $\mathcal A$, is simply ${\mathcal
A}_0$ factored by these relations. Note that the observables in
${\mathcal A}$ correspond to the correlation functions of the quantum
field $\phi$.

In the algebraic approach, a {\it state}, $\omega$, is simply a linear
map $\omega: {\mathcal A} \rightarrow {\bf C}$ that satisfies the
positivity condition $\omega (A^* A) \geq 0$ for all $A \in {\mathcal
A}$. The quantity $\omega(A)$ is interpreted as the expectation value
of the observable $A$ in the state $\omega$. 

States in the usual Hilbert space sense give rise to algebraic states
as follows: Suppose $\mathcal H$ is a Hilbert space which carries a
representation, $\pi$, of $\mathcal A$, i.e., for each $A \in
{\mathcal A}$, the quantity $\pi (A)$ is an operator on $\mathcal H$,
and the association $A \rightarrow \pi(A)$ preserves the algebraic
relations of $\mathcal A$. Let $\Psi \in {\mathcal H}$ be such that it
lies in the common domain of all operators $\pi (A)$. Then the map
$\omega: {\mathcal A} \rightarrow {\bf C}$ given by
\begin{equation}
\omega(A) = \langle \Psi | \pi (A) | \Psi \rangle
\end{equation}
defines a state on $\mathcal A$.

Conversely, given a state, $\omega$, on $\mathcal A$, we can use it to
define a (pre-)inner-product on $\mathcal A$ by
\begin{equation}
(A_1, A_2) = \omega(A_1^* A_2) \, .
\end{equation}
This may fail to define an inner product on $\mathcal A$ because,
although $(A,A) \geq 0$ for all $A \in {\mathcal A}$, there may exist
nonzero elements for which $(A,A) = 0$. However, if this happens, we
may factor the space by such zero-norm vectors. We may then complete
the resulting space to get a Hilbert space $\mathcal H$ which carries
a natural representation, $\pi$, of $\mathcal A$. The vector $\Psi \in
{\mathcal H}$ corresponding to $I \in {\mathcal A}$ then satisfies
$\omega(A) = \langle \Psi | \pi (A) | \Psi \rangle$ for all $A \in
{\mathcal A}$. The usual quantum mechanical probability rules for
determining values of the operator $\pi(A)$ in the state $\Psi$ can
then be taken over to define probability rules for the observable $A$
in the state $\omega$. We thereby obtain a complete specification of
the quantum field of a Klein-Gordon field on an arbitrary globally
hyperbolic curved spacetime insofar as the local field observables
appearing in $\mathcal A$ are concerned, i.e., in any state we can
provide the probabilities for measuring the possible values of all
observables in $\mathcal A$. No preferred notion of ``vacuum state''
or ``particles'' need be introduced, although, of course, one is free
to introduce such notions in particular spacetimes if one wishes.

As has just been seen, every state in the algebraic sense corresponds
to a state in the usual Hilbert space sense. What, then is the
advantage of formulating the theory via the algebraic approach? The
main advantage is that one is not forced to make a particular choice
of representation at the outset, i.e., one may simultaneously consider
all states arising in all Hilbert space constructions of the
theory. As a result, one may define the theory without first having to
make a choice of ``vacuum state'' or introduce any other problematical
notions. In addition, it is worth noting that the
algebraic notion of states dispenses with the unphysical states in the
Hilbert space that do not lie in the domain of the observables of the
theory; vectors in a Hilbert space representation of the theory that
do not lie in the domain of all $\pi(A)$ do not define states in the
algebraic sense.

The above provides a completely satisfactory construction of a free
Klein-Gordon field in curved spacetime insofar as observables in
$\mathcal A$ are concerned. Similar constructions can be done for all
other free (i.e., non-self-interacting) quantum fields. However, the
overall situation is still quite incomplete and unsatisfactory for at
least the following two reasons: First, even if we were only
interested in the theory of a free Klein-Gordon field, there are many
observables of interest that are not represented in $\mathcal
A$. Indeed, the observables in $\mathcal A$ are merely the $n$-point
functions of the linear field $\phi$; they do not even include
polynomial functions of $\phi$ and its derivatives (``Wick
polynomials''). A prime example of an observable of great physical
importance that is not represented in $\mathcal A$ is the
stress-energy tensor, $T_{ab}$, which would be needed to estimate
back-reaction effects of the quantum field on the spacetime
metric. Therefore, we would like to enlarge the algebra $\mathcal A$
so that it includes at least the Wick polynomials of $\phi$. Second,
we do not believe that the quantum fields occurring in nature are
described by free fields, so we would like to extend the theory to
nonlinear fields. Even in Minkowski spacetime, one understands how to
do this only at a perturbative level, but one would like to at least
generalize these perturbative rules to curved spacetime. These
perturbative rules require that one be able to define Wick polynomials
of the free field as well as time-ordered-products of polynomial
expressions in the field. Again, we need to enlarge the algebra
$\mathcal A$ so that it includes such quantities.

If a quantum field were well defined at a (sharp) spacetime event $p$,
it would be straightforward to define polynomial quantities in $\phi$
as well as time-ordered-products. However, we have already noted below
eq.(\ref{phi}) that a quantum field makes sense only as a distribution
on spacetime. Consequently, {\it a priori}, a naive attempt to define,
say, $[\phi(p)]^2$ is not likely to make any more sense than an
attempt to define the square of a Dirac delta-function. In particular,
it would be natural to attempt to define the smeared Wick power
$\phi^2 (f)$ by a formula like
\begin{equation}
\phi^2 (f) = \lim_{n \rightarrow \infty} 
\int \phi(x) \phi(y) f(x) F_n (x,y) d^4 x d^4 y
\label{phi2}
\end{equation}
where $F_n(x,y)$ is a sequence of smooth functions that approaches the
Dirac delta-function $\delta(x,y)$. However, the right hand side
diverges in the limit, so some sort of ``regularization'' of this
expression must first be done in order to make the limit well behaved.

Once the Wick powers $\phi^k (f)$ have been defined, it would be an
easy matter to define the time-ordered-product $T(\phi^{k_1} (f_1)
\dots \phi^{k_n} (f_n))$ by a straightforward ``time ordering'' of the
factors in the case where the supports of $f_1, \dots, f_n$
have suitable causal properties so that they can be put in a well
defined time order. Indeed, using induction in the number, $n$, of
factors, it is straightforward to define $T(\phi^{k_1} (f_1) \dots
\phi^{k_n} (f_n))$ whenever the intersection of the supports of $f_1,
\dots, f_n$ vanishes. However, it is not straightforward to extend
this distribution to the ``total diagonal'', i.e., to the case where
the supports of $f_1, \dots, f_n$ have nonvanishing mutual
intersection.

From the way I have described the regularization issues above, it
might seem that the most difficult problem would be to define Wick
polynomials and that the definition of time-ordered-products would be
a minor addendum to this problem. In fact, however, in Minkowski
spacetime Wick polynomials are easily defined by a ``normal ordering''
prescription, which can be interpreted as subtracting off the vacuum
expectation value of the field quantities before taking the kind of
limit appearing on the right side of eq.(\ref{phi2}). On the other
hand, the problem of extending time-ordered-products to the total
diagonal corresponds to the problem of renormalizing all Feynman
diagrams---an extremely difficult and complex problem.

There are major issues of principle that must be overcome in order to
extend the Minkowski spacetime regularization and renormalization
prescriptions to curved spacetime. The normal ordering prescription
for defining Wick polynomials in Minkowski spacetime relies on the
existence of a preferred vacuum state, with respect to which the
``normal ordering'' is carried out. However, we have already seen that
in a general curved spacetime, there does not appear to exist any
notion of a preferred vacuum state. Furthermore, the renormalization
prescriptions used to define time-ordered-products in Minkowski
spacetime make use of ``momentum space methods'' (i.e., global Fourier
transforms of quantities) and/or ``Euclidean methods'' (i.e., analytic
continuation of expressions defined on Euclidean space rather than
Minkowski spacetime). These methods, in turn, require Poincare
symmetry, the existence of a preferred, Poincare invariant vacuum
state, and/or the ability to ``Euclideanize'' Minkowski spacetime by
the transformation $t \rightarrow it$. All of these features are
absent in a general, curved spacetime.

It was already understood by the late 1970's that it should be
possible to define the stress-energy tensor, $T_{ab}$, of a quantum
field $\phi$ only on a restricted class of states, namely the
so-called Hadamard states, $\omega_H$, whose two-point distribution
$\omega_H (\phi(x), \phi(y))$ has a short distance singularity
structure as $y \rightarrow x$ of a particular form (see, e.g.,
\cite{w2}). For Hadamard states, a prescription for defining the
expectation value, $\omega_H (T_{ab})$, can be given that involves the
subtraction from $\omega_H (\phi(x), \phi(y))$ of a locally and
covariantly constructed Hadamard parametrix rather than a vacuum
expectation value \cite{w2}. The resulting prescription defines $\omega_H
(T_{ab})$ in an entirely satisfactory manner that does not require a
choice of vacuum state. Indeed, this prescription is local and
covariant in the sense that the value of $\omega_H (T_{ab})$ at a
point $p$ depends only on the spacetime geometry and the behavior of
$\omega_H$ in an arbitrarily small neighborhood of $p$. It is not
difficult to show that, even if one could make a unique choice of
vacuum state in all spacetimes, normal ordering would not provide a local
and covariant definition of $\omega_H (T_{ab})$.

However, although the above prescription provides a satisfactory
definition of the expectation value of the stress-energy tensor in
Hadamard states and can be generalized to define the expectation value
of higher Wick powers, it does not define $T_{ab}$ or other Wick
powers as an element of an enlarged algebra. Indeed, in a Hilbert
space representation of the theory, the above prescription would
merely define $T_{ab}$ as a quadratic form on Hadamard states rather
than as an operator-valued-distribution, so no probability rules for
measuring the possible values of $T_{ab}$ would be
available. Furthermore, it is worth mentioning that the
characterization of ``Hadamard states'' in terms of their
short-distance singularity structure is extremely cumbersome to work
with. Finally, by the mid-1990's it was still very far from clear how
to perform the much more difficult and complex renormalizations that
would be needed to define time-ordered-products in curved spacetime.

\section{Progress Since the Mid-1990's}

During the past decade, the algebra of observables for a free quantum
field has been extended to include all Wick polynomials and
time-ordered-products, so that, in particular, the perturbative
renormalization of interacting quantum fields in curved spacetime is
now rigorously well defined. Much of this progress has resulted from
the importation of methods of ``microlocal analysis'' into the
theory. In essence, microlocal analysis provides a refined
characterization of the singularities of a distribution. If one has a
distribution, $\alpha$, defined in a neighborhood of point $p$ on a
manifold $M$, one can multiply $\alpha$ by a smooth function $f$ with
support in an arbitrarily small neighborhood of $p$, such that $f(p)
\neq 0$. One can then examine the decay properties of the Fourier
transform of $f \alpha$ (where the Fourier transform can be defined by
choosing an arbitrary embedding of a neighborhood of the support of $f
\alpha$ into Euclidean space). If $\alpha$ were smooth in a
neighborhood of $p$, then, for $f$ with support in this neighborhood,
the Fourier transform of $f \alpha$ would decay rapidly in all
directions in Fourier transform space, $k$, as $|k| \rightarrow
\infty$. Thus, the failure of the Fourier transform of $f \alpha$ to
decay rapidly characterizes the singular behavior of $\alpha$ at
$p$. If, for all choices of $f$, the Fourier transform of $f \alpha$
does not decay rapidly in a neighborhood of the direction $k$, then
one says that the pair $(p,k)$ lies in the {\it wavefront set}, ${\rm
WF}(\alpha)$, of $\alpha$. One can naturally identify ${\rm
WF}(\alpha)$ with a subset of the cotangent bundle of the manifold,
$M$. The wavefront set thereby provides a characterization of not only
the {\it points} in $M$ at which $\alpha$ is singular, but also the
{\it directions} (in the cotangent space) at which it is
singular. This refined characterization of the singularities of
distributions can enable one to define operations that normally are
ill defined. For example, if $\alpha$ and $\beta$ are distributions,
then it normally will not make mathematical sense to take their
product. However, if it is the case that whenever $(p,k) \in {\rm
WF}(\alpha)$, we have that $(p,-k) \notin {\rm WF}(\beta)$, then the
product $\alpha \beta$ can be defined in a natural way via the Fourier
convolution formula.

By providing rules for, e.g., when products of distributions are well
defined as distributions, microlocal analysis provides an extremely
useful calculus for determining whether proposed
regularization/renormalization schemes are well defined. Since the
analysis is completely local in nature, it provides an ideal tool for
analyzing the behavior of local field observables.

The first significant application of microlocal analysis to quantum
field theory in curved spacetime occurred in the Ph.D. thesis of
Radzikowski \cite{r}, a student of Wightman. Radzikowski was concerned
with proving a conjecture, due to Bernard Kay, that stated that if a
quantum state had a two-point function whose short-distance
singularities are of the Hadamard form, then it could not have any
additional singularities at large spacelike separations (``local
Hadamard form implies global Hadamard form''). Radzikowski employed
the tools of microlocal analysis to prove this
conjecture.  In particular, in the course of his analysis, he proved
that the (quite cumbersome) characterization of Hadamard states in
terms of the detailed local singularity structure of $\omega_H
(\phi(x), \phi(y))$ is equivalent to a very simple condition on the
wavefront set of this distribution, namely ${\rm WF} [\omega_H
(\phi(x), \phi(y))]$ is the subset of the cotangent bundle of $M
\times M$ consisting of all points $(x,y;k,l)$ such that $x$ and $y$
are connected by a null geodesic $\gamma$ with future-directed tangent
$k^a = g^{ab} k_b$ at $x$ and with $l_a$ being minus the parallel
transport along $\gamma$ of $k_a$ to $y$.

It is worth mentioning that there was an interesting historical
interplay between microlocal analysis and quantum field theory in
curved spacetime. In the late 1960's Hormander visited the Institute
for Advanced Study in Princeton and interacted with Wightman. Wightman
explained to Hormander what the ``Feynman propagator'' is in Minkowski
spacetime, and a characterization of ``Feynman parametrices'' in a
general curved spacetime in terms of wavefront set properties can be
found in the classic paper of Duistermaat and Hormander
\cite{dh}. Conversely, Wightman realized that the methods of
microlocal analysis could potentially be useful in the formulation of
quantum field theory in curved spacetime. For example, in de Sitter
spacetime, there is no globally timelike Killing field and therefore
no global notion of energy that is positive. Therefore, it does not
appear that one could impose a global spectral condition on a quantum
field analogous to the requirement of positivity of energy in the
Minkowski case. However, one might be able to impose a ``microlocal
spectral condition'' on the local quantum field observables. Shortly
after his interactions with Hormander, Wightman had a student,
S. Fulling, who was interested in quantum field theory in curved
spacetime, and he suggested to Fulling that he investigate the
possible application of microlocal analysis to quantum field theory in
curved spacetime. However, after spending some effort in studying
microlocal analysis, Fulling decided that his efforts would be better
spent on other projects. Among the other projects that Fulling then
investigated in his thesis was the inequivalence of different
quantization schemes. In particular, he showed that quantization in
the ``Rindler wedge'' of Minkowski spacetime using a Lorentz boost
Killing field to define a notion of ``time translations'' gave rise to a
different notion of ``vacuum state'' than the restriction of the usual
Minkowski vacuum to this region. This work provided the mathematical
basis for Unruh's subsequent analysis discussed above
\cite{u2}. However, Wightman had to wait another twenty years before
he had another student interested in quantum field theory in curved
spacetime. When Radzikowski began to apply the methods of microlocal
analysis to analyze Kay's conjecture, Wightman was well prepared to provide
plenty of encouragement.

After Radzikowski's work, it became clear to Fredenhagen and
collaborators that microlocal analysis should provide the needed tools
for analyzing the divergences occurring in quantum field theory in
curved spacetime. Brunetti, Fredenhagen, and Kohler \cite{bfk} showed
that if one considers a Fock representation associated with an
arbitrary Hadamard vacuum state $\omega_0$, then normal ordering can
be used to define Wick polynomials as operator-valued-distributions on
this Hilbert space. Indeed, a larger algebra, $\mathcal W$, of field
observables---large enough to include all time-ordered-products---can
be defined in this manner. Brunetti and Fredenhagen \cite{bf} also
formulated a microlocal spectral condition that should be imposed on
time-ordered-products. However, as previously mentioned, a normal
ordering prescription cannot yield a local and covariant definition of
Wick polynomials. Furthermore, the construction of $\mathcal W$ given
in \cite{bfk} invokes an arbitrary choice of Hadamard vacuum state
$\omega_0$. Nevertheless, it can be shown that, as an abstract
algebra, $\mathcal W$ does not depend on the choice of $\omega_0$, so
it is a legitimate candidate for the desired enlarged algebra of
observables. Thus, the key remaining issue was to determine which
elements of $\mathcal W$ properly represent the ``true'' Wick
polynomials and time-ordered-products.

A key condition to be imposed on the definition of Wick polynomials
and time-ordered-products is that they be local and covariant fields.
As mentioned in the previous section, this condition had been imposed
on the definition of the expectation value of the stress-energy.
However, the formulation of this notion given in \cite{w2} was not
adequate for the present purpose, and a more general formulation had
to be given \cite{hw1,bfv}.

With these key ideas and constructions in place, it was possible to
prove the following results \cite{hw1}, \cite{hw2}-\cite{hw4}: (1)
There exists a well defined prescription for defining all Wick
polynomials that is local and covariant and satisfies a list of
additional reasonable properties, including appropriate scaling
behavior and continuous/analytic variation under continuous/analytic
changes in the metric \cite{hw1}. This prescription is unique up to certain
``local curvature ambiguities''. For example, for a Klein-Gordon
field, $\phi$, the prescription for $\phi^2$ is unique up to
\begin{equation}
\phi^2 \rightarrow \phi^2 + (c_1 R + c_2 m^2) I 
\end{equation}
where $c_1, c_2$ are arbitrary constants, $R$ denotes the scalar
curvature, and $I$ denotes the identity element of $\mathcal W$. For a
massless field in Minkowski spacetime, all of the ambiguities
disappear, and the prescription agrees with normal ordering with
respect to the usual Minkowski vacuum state. However, on a general
curved spacetime, the prescription for defining $\phi^2$ and other
Wick polynomials does not agree with normal ordering with respect to
any choice of vacuum state. (2) There exists a prescription for
defining all time-ordered-products that is local and covariant, that
satisfies the microlocal spectral conditions \cite{bf}, and that
satisfies a list of additional reasonable properties \cite{hw2}. This
prescription is unique up to ``renormalization ambiguities'' of the
type expected from Minkowski spacetime analyses, but with additional
local curvature ambiguities. (3) Theories that are renormalizable in
Minkowski spacetime remain renormalizable in curved spacetime. For
renormalizable theories, renormalization group flow can be defined in
terms of the behavior of the quantum field theory under scaling of the
spacetime metric, $g_{ab} \rightarrow \lambda^2 g_{ab}$, \cite{hw3}.
(4) Additional renormalization conditions on time-ordered-products can
be imposed so that, order by order in perturbation theory, for an
arbitrary (not necessarily renormalizable) interaction, (i) the
interacting field satisfies the classical interacting equation of
motion and (ii) the stress-energy tensor of the interacting field is
conserved \cite{hw4}. All of the above results have been obtained
without any appeal to a notion of ``vacuum'' or ``particles''.

These and other results of the past decade have demonstrated that
quantum field theory in curved spacetime has a mathematical structure
that is comparable in depth to such theories as classical general
relativity. In particular, it is highly nontrivial that quantum field
theory in curved spacetime appears to be mathematically
consistent. Although quantum field theory in curved spacetime cannot
be a fundamental description of nature since gravity itself is treated
classically, it seems hard to believe that it is not capturing some
fundamental properties of nature.

The above results suffice to define interacting quantum field theory
in curved spacetime at a perturbative level. However, it remains very
much an open issue as to how to provide a non-perturbative formulation
of interacting quantum field theory in curved spacetime. It is my hope
that significant progress will be made on this issue in the coming years.

\bigskip 
\begin{center}
{\bf Acknowledgements} 
\end{center}
This research was supported in part by NSF grant PHY-0456619 to the
University of Chicago.

\end{document}